\documentclass[12pt,letterpaper]{article}

\usepackage{graphicx}
\usepackage[body={17.5cm, 22cm},right=2.2cm]{geometry}
\usepackage{amssymb}
\usepackage{amsmath} 

\newcommand\ee{\end{equation}}
\newcommand\be{\begin{equation}}
\newcommand\eea{\end{eqnarray}}
\newcommand\bea{\begin{eqnarray}}
\newcommand{\sfrac}[2]{{\textstyle\frac{#1}{#2}}}

\begin{document}

\begin{center}

~

\vspace{1.cm}

{\LARGE \bf{
On Super-Planckian Fields \\[.2cm]
at Sub-Planckian Energies
}}\\[1cm]
{\large Alberto Nicolis}
\\[0.6cm]

{\small \textit{
Department of Physics and ISCAP, \\Columbia University,
New York, NY 10027, USA }}

\end{center}

\vspace{.8cm}
\begin{abstract}
\noindent
For a light scalar coupled to gravity, I study the gravitational backreaction associated with large field variations. 
I show a generic obstruction in sourcing  a super-Planckian scalar profile without making the whole experiment collapse into a black hole.
In empty space the scalar variation obeys an absolute bound, of order of the Planck scale.
A Newtonian analysis suggests that {\em inside} its sources the scalar can undergo arbitrarily large variations without causing large gravitational backreactions. However the maximum attainable $\Delta\varphi$ increases only logarithmically with the size of the source.
The bound straightforwardly generalizes to any number of dimensions, and to moduli space-like cases, where it applies to the invariant length in field space as measured by the kinetic metric.
\end{abstract}

\section{Introduction}

It appears that string theory abhors super-Planckian scalar fields. 
In all known string-theory vacuum constructions, the geodesic distance in moduli space between any two points is finite, and of order of the Planck scale to the appropriate power. It diverges logarithmically along certain directions, but only at the expense of making an infinite tower of states become massless, thus in practice causing the effective field theory to break down \cite{swampland1, swampland2}. In particular, there have been several attempts to construct string-theoretical inflationary models that feature super-Planckian excursions for the inflaton $\varphi$, but this has proved to be very difficult  (see e.g.~ref.~\cite{banks}, and ref.~\cite{liam} for a recent review). This is an observationally relevant issue, since a super-Planckian $\Delta \varphi$ during inflation is necessary for primordial tensor modes to be observable
in CMB experiments \cite{lyth}.

On the other hand, from a purely low-energy, field-theoretical viewpoint a super-Planckian vev for a scalar $\varphi$ does not necessarily lead to a breakdown of the low-energy effective theory. If the potential for $\varphi$ is sufficiently flat and gradients of $\varphi$ are small, Planckian values of $\varphi$ do not correspond to Planckian energy densities. Also, the flatness of the potential can be protected against quantum corrections coming for instance from graviton loops, by an approximate shift-symmetry on $\varphi$, thanks to which non-renormalizable operators of the form $\varphi^n/M_{\rm Pl}^{n-4}$ in the effective potential are further suppressed.
It is commonly believed that quantum gravity breaks all continuous global symmetries, in which case  the shift symmetry above could not be used to suppress non-renormalizable operators beyond the expected $M_{\rm Pl}$-suppression. (The breaking of the shift symmetry might even only show up non-perturbatively, through terms $\sim \exp(- 1/ G \varphi^2)$, which would however become important for $\varphi$ values of order of the Planck scale or larger.)
However this obstacle in principle can be circumvented---like in the example of ref.~\cite{paolo}, where the (approximate) shift symmetry on $\varphi$ is a consequence of a gauge symmetry, which is unbroken by quantum gravitational effects.

It appears then that the absence of super-Planckian scalars is a recurrent pattern in string theory whose necessity is not obvious from the low-energy viewpoint.
It may be that consistently coupling field theories to quantum gravity inevitably forbids super-Planckian scalar fields. This would be similar to other conjectured properties of quantum gravity \cite{swampland1, weakgravity, swampland2}, which have no obvious explanations in purely low-energy terms. Or it may be that models with super-Planckian fields are sporadically realized in the string landscape, and so far we have just been unlucky.

In this note I assume that such super-Planckian scalars exist, and I question whether an observer in asymptotically flat space can actually set up a local experiment that allows him or her to probe those remote field values. We will see that there is a concrete tension between setting up a field-profile that spans a huge field range, and keeping the gravitational backreaction small enough not to make the whole experiment collapse into a black hole.
In the following I will only consider massless fields. The presence of any (positive definite) potential can only sharpen the tension: for fixed field profile, it would add a positive contribution to the local energy density, thus strengthening the gravitational backreaction.

\section{The Newtonian bound}

Consider a canonically normalized scalar $\varphi$ minimally coupled to gravity \footnote{We are using the $(-,+,+,+)$ signature for the metric.
},
\be \label{phiaction}
S_{\varphi} = - \int \! d^3x \sqrt{-g} \: \sfrac12 (\partial \varphi)^2 \; .
\ee
Suppose we want to set up an actual experiment that allows us to measure a huge excursion for $\varphi$ between two points, say $|\vec x| = \infty$ and the origin $\vec x= 0$. We need to properly arrange sources for $\varphi$ that produce the profile we want. We will not consider the sources' internal dynamics---we will just assume that we can arrange sources in any way we want, and that their gravitational backreaction is negligible. In this sense our estimates will be conservative, in that requiring for self-consistent source dynamics and taking into account their gravitational field will generically give stronger constraints.

For simplicity we will consider static configurations $\varphi(\vec x)$, and we will assume that the gravitational backreaction is everywhere small so that we can use Newtonian gravity (we will relax the latter assumption in sect.~\ref{GRcase}). Then the Newtonian potential $\Phi$ solves Poisson's equation,
\be
\nabla^2 \Phi = 4\pi G \: \sfrac12  |\vec  \nabla \varphi |^2 \; .
\ee
Our goal is maximize $\Delta \varphi$ between the origin and infinity while keeping $\Phi$ small. Without loss of generality, we choose vanishing boundary conditions at infinity for $\varphi$.
The Newtonian potential at the origin is
\be
\Phi ( 0 ) = - \sfrac12 G \, \int \! d^3 x \: \frac{ |\vec  \nabla \varphi |^2}{|\vec x|} \; ,
\ee
and although generically we would like to impose that $\Phi$ is small everywhere in space, we will see that requiring it be small at the origin is enough. Also, since $\Phi$ is negative definite in the following by `Newtonian potential' we will refer to $|\Phi|$.

First, notice that for maximizing the field excursion while keeping $\Phi(0)$ small it is best to set up spherically symmetric configurations. To see this, expand a generic $\varphi$ configuration in spherical harmonics around the origin
\be
\varphi(\vec x) = \sum_{\ell,m} Y^\ell _m (\theta, \phi) \, \varphi_{\ell m} (r) \; ,
\ee
and notice that each multipole gives a positive definite contribution to the Newtonian potential,
\be
 | \Phi (0) | = \sfrac12 G \sum _{\ell, m} \int_0 ^\infty dr \bigg[ r \big( \partial_r \varphi_{\ell m} \big)^2
 + \frac{\ell(\ell+1)}{r}  \,  \varphi_{\ell m}^2 \bigg] \; .
\ee
Then, we can lower the Newtonian potential if we are able to discard some of the multipoles while keeping the overall $\varphi$ excursion between the origin and infinity fixed. We can in fact keep just the {\em monopole} and discard all the others: the monopole vanishes at infinity, since by linear independence of the spherical harmonics the vanishing boundary condition for $\varphi$ implies that all $\varphi_{\ell m}$'s vanish at $r=\infty$. Also since $\varphi$ is single-valued at the origin, all multipoles with $\ell \ge 1$ vanish there. Thus the monopole in enough to reproduce the overall $\Delta \varphi$.
Therefore for a given field excursion spherical configurations allow for smaller $\Phi$'s, or equivalenty, for a given $\Phi$ they allow for larger field excursions. We will therefore concentrate on radial profiles $\varphi(r)$.
Also notice that, given the positivity of $(\vec \nabla \varphi)^2$, for spherically symmetric $\varphi$'s the maximum of the Newtonian potential is achieved at the origin. In spherical coordinates we have
\be \label{sphericalPhi}
\Phi_{\rm max} \equiv |\Phi (0) | = 2 \pi G \int _0 ^{\infty} \! dr \: r \big( \varphi'(r) \big) ^2 \; . 
\ee

Next, note that placing just a compact source at the origin will not give us a super-Planckian excursion {\em outside} the source. Indeed the field outside the source is
\be
\varphi(r) = \varphi_* \frac{r_*}{r} \; ,
\ee
where $\varphi_*$ is the field value at the source boundary, and $r_*$ is the source radius. On the other hand, we can get a lower bound on the Newtonian potential at the origin by truncating the integral in eq.~(\ref{sphericalPhi}) at the source's surface $r = r_*$, keeping only the contribution coming from outside the source. We get
\be
\Phi_{\rm max} \ge \pi G \, \varphi_* ^2 \; ,
\ee
which is independent of the source size. 
(The inequality is saturated if the field profile flattens out immediately below the source's surface, $\varphi = \varphi_*$ for $r < r_*$, thus giving no extra contribution to the Newtonian potential.)
The overall field variation $(\Delta \varphi)_{\rm out}$ between $r = \infty$ and the source's surface is $\varphi_*$ itself. 
Therefore
\be \label{outside}
G \cdot (\Delta \varphi)_{\rm out}^2 \le \frac{ \Phi_{\rm max} }{\pi} \ll 1 \; ,
\ee
where we imposed that the Newtonian potential is small everywhere.

We are thus led to consider the field-profile {\em inside} the source, for $r < r_*$.  Notice that we want $r_*$ to be finite: any local experiment we can perform in a lab will have a finite size. We will see that the finiteness of $r_*$ is important. 
For simplicity we will neglect the contributions from outside the source we just computed, as they do not lead to super-Planckian excursions.
Then the Newtonian potential at the origin, eq.~(\ref{sphericalPhi}), is
\be \label{functional}
\Phi_{\rm max} = 2 \pi G \int _0 ^{r_*} \! dr \: r \big( \varphi'(r) \big) ^2 \; . 
\ee
We want to maximize
\be \label{delta}
(\Delta \varphi)_{\rm in} = -\int_0 ^{r_*} \varphi'(r) dr
\ee
by varying the profile $\varphi(r)$,
while keeping $\Phi_{\rm max}$ fixed, and small. It is clear that a completely equivalent variational problem is to minimize $\Phi_{\rm max}$ while keeping $(\Delta \varphi)_{\rm in} $ fixed. Indeed, in the vector space of field profiles both problems correspond to finding the points at which the $\Phi_{\rm max} = {\rm const}$ ellipsoids are tangent to the $(\Delta \varphi)_{\rm in}  = {\rm const}$ hyperplanes.
But keeping $(\Delta \varphi)_{\rm in} $ fixed is the same as keeping the boundary conditions for $\varphi$ fixed, so that the minimum of $\Phi_{\rm max}$ should be attained by the solution to the Euler-Lagrange equation,
\be \label{logr}
\frac{d}{dr} \big ( r \, \varphi'(r) \big) = 0 \qquad \Rightarrow \qquad \varphi(r) = a \log r + b \; ,
\ee 
where $a$ and $b$ are integration constants.
We see that $\varphi(r)$ diverges logarithmically at the origin, and so does $\Phi_{\rm max}$ if we plug back the solution for $\varphi$ into eq.~(\ref{functional}), while instead we wanted to keep $\Phi_{\rm max}$ small. Technically this is happening because eq.~(\ref{functional}) as a functional of $\varphi(r)$ is not {\em coercive}---the coefficient of ${\varphi'} ^2$ goes to zero at the origin. As a consequence, the existence of a minimum with fixed, finite boundary condition is not guaranteed. However this is physically irrelevant. Any experiment will have a finite resolution in space: we will not be able to resolve distances below some UV cutoff $\ell_{\rm UV}$, ultimately the Planck length $\ell _{\rm Pl}$. Perhaps more importantly, a field profile like (\ref{logr}) has infinite energy density at the origin, thereby producing super-Planckian curvatures. Then it is physically very sensible to cutoff both integrals in eq.~(\ref{functional}, \ref{delta}) at $\ell_{\rm UV}$ rather than extend them all the way down to the origin, by assuming that below $\ell_{\rm UV}$ the $\varphi$-profile flattens out. If we do so, the Euler-Lagrange equation and its solution are still the same and we get
\be
(\Delta \varphi)_{\rm in}  = a \log (r_*/\ell_{\rm UV}) \; , \qquad  \Phi_{\rm max} = 2 \pi G \, a^2 \log (r_*/\ell_{\rm UV}) \; ,
\ee
where $a$ is a free parameter. For a given  $ \Phi_{\rm max} \ll 1$ we thus have:
\be \label{bound}
G \cdot (\Delta \varphi)_{\rm in} ^2 = \frac {\Phi_{\rm max} }{2 \pi} \,  \log (r_*/\ell_{\rm UV}) \; .
\ee
We see that for fixed Netwonian potential the optimal $(\Delta \varphi)_{\rm in} ^2$ grows logarithmically with the source's size $r_*$. Then it appears that it is possible  to source and measure arbitrarily large field values without making the lab collapse into a black hole, but only at the expense of building exponentially large experiments.


\section{Generalizations}

\subsection{Non-canonical, multi-field case}
The above results trivially generalize to the case of a non-canonically normalized field $\psi$ with Lagrangian ${\cal L} = -\frac12 G( \psi) (\partial \psi)^2$. It is always possible to go to canonical normalization,
\be
\varphi \equiv \int \! \sqrt{G(\psi)} \: d\psi \; ,
\ee
so that eqs.~(\ref{outside}, \ref{bound}) apply to the invariant length in field space as measured by the metric $G(\psi)$.

The situation is more complicated for a multi-field case with Lagrangian
\be \label{multi}
{\cal L} = - \sfrac12 G_{AB}( \psi) \, \partial_\mu \psi^A \partial^\mu \psi^B \; ,
\ee
because now the length in field space depends on the path, whereas in the single field case there exists a globally defined, single-valued $\varphi$ that we can use as an invariant length. So, for instance, if for a given static configuration of the $\psi^A$'s we now take a path $\gamma$ that starts at $|\vec x| = \infty$ and lingers in complicated zigzags in field-space in its way to the origin of real space, we can make the invariant field-space length
\be
L(\gamma) = \int _\gamma  \sqrt{G_{AB}(\psi) \, d \psi^A d \psi^B}
\ee
as long as we like. However, this is just the statement that to get a unique answer the distance between two points has to be measured along geodesics. Here we have two choices, since we have two spaces. We can choose either the real-space geodesics---straight lines in our Newtonian approximation---or the field-space geodesics associated with the metric $G_{AB}$. The latter choice seems to be the more appropriate to characterize the geometry of field-space. The former however is simpler to work with, and leads directly to our bounds eqs.~(\ref{outside}, \ref{bound}), as we now show.

Consider indeed in real-space all the rays emanating from the origin and going out  to infinity. Each ray $\gamma$ corresponds to some direction $(\theta, \phi)$, and is parameterized by $r$. Then, for a given static configuration of the $\psi^A$'s, with each ray $\gamma(\theta, \phi)$ there is an associated field-space length,
\be
L(\theta, \phi) =  \int _{\gamma(\theta, \phi)}  \sqrt{G_{AB}(\psi) \, d \psi^A d \psi^B} =
			  \int_0 ^\infty \! dr \sqrt{G_{AB}(\psi) \, \partial_r \psi^A \partial_r \psi^B} \, .
\ee
We can also make this a function of $r$ by truncating the integral at finite $r$,
\be
L(r, \theta, \phi) \equiv \int_r ^\infty \! dr' \sqrt{G_{AB}(\psi) \, \partial_{r'} \psi^A \partial_{r'} \psi^B} \,
\ee
This defines a scalar field throughout space with the possible exception of the origin, where $L$ might not be single valued. $L(\vec x)$  is nothing but the length of the trajectory traveled in field space in moving from $r=\infty$ to finite $r$  along the $(\theta,\phi)$ direction; as we will see, $L(\vec x)$ plays the role of our canonically normalized $\varphi$ above.

The Newtonian potential is sourced by the fields' energy density
\be \label{energy}
\rho = \sfrac 12 G_{AB} \, \big( \vec \nabla \psi^A \cdot \vec \nabla \psi^B \big) \ge \sfrac 12 G_{AB} \, \partial_r \psi^A \partial_r
\psi^B = \sfrac12 \, (\partial_r L)^2 \; ,
\ee
where obviously we assumed that $G_{AB}$ is a positive-definite metric. The equality only holds for spherically symmetric configurations $\psi^A(r)$. Then
at the origin we have
\be \label{spherical}
|\Phi (0)| = G \int \! d^3 x \: \frac{\rho}{|\vec x|} \ge
 \sfrac12 G \int \! d^3 x \: \frac{( \partial_r L )^2}{r}
 =
 \sfrac12 G \sum _{\ell, m} \int_0 ^\infty \! dr \:  r \big( \partial_r L_{\ell m} (r) \big)^2 \; ,
\ee
where we expanded $L(\vec x)$ in spherical harmonics, $L(\vec x) = \sum Y^\ell _m(\theta, \phi) L_{\ell m} (r)$.
Once again each multipole of $L$ gives a positive-definite contribution to eq.~(\ref{spherical}).
However now the monopole is generically not enough to reproduce the maximum possible length, because $L(\vec x)$ itself is not single-valued at the origin, and so for a generic field configuration several multipoles will be important in attaining the maximum of $L$ along some particular direction $(\theta_0, \phi_0)$. 
Therefore eqs.~(\ref{energy}, \ref{spherical}) suggest that spherical configurations are the most efficient, in that at  least they saturate the inequalities, but in this multi-field case we lack a complete proof.

%

We are therefore led once again to consider radial profiles $\psi^A(r)$. They saturate the inequalities in eqs.~(\ref{energy}, \ref{spherical});
then our arguments that yielded eq.~(\ref{bound}) apply unaltered, apart from the typographical replacement of $\varphi(r)$ with $L(r)$. 
In conclusion, in a generic multi-field case we get a bound on the invariant length $L$ in field space as measured along straight-lines in real space,
\be
G \cdot L_{\rm in}^2 \le \frac {\Phi_{\rm max}}{2 \pi} \,  \log (r_*/\ell_{\rm UV}) \; .
\ee

\subsection{Generic $D$}

In $D$ space-time dimensions we have
\be
[G] = {\rm length}^{D-2} \; , \qquad [\varphi] = {\rm length}^{-\frac{D-2}2}
\ee
and so bounds like eqs.~(\ref{outside}, \ref{bound}) are still dimensionally correct.
Also the arguments for the efficiency of spherically symmetric configurations apply to generic $D$, and in the case of a radial profile $\varphi(r)$ we have
\be
\Phi_{\rm max} \sim G \int \! d^{D-1} x \: \frac{|\vec \nabla \varphi|^2}{|\vec x|^{D-3}} \sim G \int \! dr \: r \big( \varphi'(r) \big) ^2 \; ,
\ee
where we are dropping $\pi$'s etc.
We see that the variational problem is the same as in $D=4$, thus leading to the same bound, eq.~(\ref{bound}), with the same logarithmic enhancement.


\section{The bound in General Relativity}\label{GRcase}

We now move to study the same problem in General Relativity, relaxing the assumption that the Newtonian potential is small everywhere. For simplicity we stick to static, spherically symmetric configurations for the scalar and the metric. Also we require that the metric be asymptotically flat. We assume once again that we have a compact spherical source for the scalar sitting at the origin, and we ask what is the maximum scalar field excursion we can measure in moving from infinity to the source's surface. We will comment on what may happen inside the source in sect.~\ref{discussion}.

In the case of a minimally coupled, massless scalar the general solution was found by Buchdahl \cite{buchdahl}. Starting from any static {\em vacuum} solution of Einstein's equation, it is possible to explicitly construct a new solution with the same symmetries, where the scalar is turned on  \cite{buchdahl}. In our case the relevant vacuum solution is obviously the Schwarzschild metric. Calling $f(r)$ the usual  Schwarzschild function
\be
f(r) \equiv 1- \frac{2G m}{r}  \; ,
\ee
the corresponding solution in the presence of the scalar is \cite{buchdahl}
\begin{eqnarray}
ds^2 & = & -f (r)  ^ \beta dt^2 + f (r) ^{-\beta}  dr^2 +  r^2 f(r)^{1-\beta} d\Omega^2 \; , \label{metric}\\
\varphi(r) & = & \varphi_0 \, \log 1/f(r) \; ,  \label{phi}
\end{eqnarray}
where $\varphi_0$ is a free parameter and
\be \label{beta}
\beta \equiv \sqrt{1 - 16 \pi G \, \varphi_0 ^2} \; .
\ee
We therefore have a two-parameter  family of solutions, the two independent parameters being $\varphi_0$ and $m$. This exhausts all possible solutions with the desired properties---staticity, spherical  symmetry, and asymptotic flatness---as shown
 in ref.~\cite{wyman}.

A few comments are in order:
\begin{itemize}
\item[{\em (i)}]
From eq.~(\ref{beta}) we see that {\em there is simply no real solution for $\varphi_0^2 > 1/(16 \pi G)$}. However this is not enough yet to bound the overall excursion of $\varphi$, because of the logarithmic factor in eq.~(\ref{phi}).
Indeed $f(r)$ goes to zero at $r=  2Gm$, thus making $\varphi$ diverge if that point is accessible, or indefinitely approchable.

\item[{\em (ii)}]
Unlike in the Schwarzschild solution, here the singularity at $r = 2Gm$ is physical, as can be seen by computing curvature invariants, like  e.g.~the Ricci curvature,
\be
{\cal R} = 2 (1-\beta^2) \, G^2m^2 \cdot \frac{1}{r^4  f(r)^{2 -\beta} } \; .
\ee
As $0 \le \beta <1$ (see eq.~(\ref{beta})), ${\cal R}$ diverges when $f(r) \to 0$. 
This is just a manifestation of the general `no-hair' statement---that there are no black holes solutions with nontrivial scalar profiles, because the scalar's stress-energy tensor would blow up at the horizon \cite{nohair}.
For us this simply means that for a regular solution to exist at all, the source should enter the picture before we hit the singularity. That is, the interior solution---which we don't know---should be glued to eq.~(\ref{metric}) at some $r_* > 2Gm$, exactly like for an ordinary star the would-be Schwarzschild radius is inside the star.

\item[{\em (iii)}]
By direct inspection of  the metric's asymptotic behavior as $r \to \infty$,
\be
ds^2 \simeq - \left(1- \beta \frac{2 G m}{r} \right) dt^2 + \left(1+ \beta \frac{2 G m}{r} \right) dr^2 + r^2 
\left(1-(1- \beta) \frac{2 G m}{r} \right) d\Omega^2 
 \; , 
\ee
we get that the total ADM mass is {\em not} the mass parameter $m$ appearing in $f(r)$. Rather it is
\be \label{mass}
M_{\rm ADM} = \beta m \; .
\ee

\end{itemize}

\noindent
We now have all the ingredients to find an absolute bound on the overall excursion $\Delta \varphi$ bewteen $r = \infty$ and the source's surface at $r=r_*$,
\be \label{phistar}
(\Delta \varphi)_{\rm out} = \varphi_* \equiv \varphi_0 \log 1 / f(r_*) \; .
\ee
$(\Delta \varphi)_{\rm out}$ can be made large at will if, for fixed $\varphi_0$, the source's radius $r_*$ can be made arbitrarily close to the singularity at $r=2Gm$. However this is not possible. To see why, consider the total mass of the configuration, eq.~(\ref{mass}). For generic asymptotically flat spacetimes, we just know that whenever the dominant energy condition is obeyed $M_{\rm ADM}$ is non-negative, and vanishes exactly only for flat space \cite{positiveenergy1, positiveenergy2}.
However for static, spherically symmetric configurations we can say much more. Indeed, using coordinates such that the metric is
\be
ds^2 = -B(R) \, dt^2 + A(R) \, dR^2 + R^2 d\Omega^2 \; ,
\ee
the total mass can be expressed as (see e.g.~ref.~\cite{weinberg})
\be \label{integralrho}
M_{\rm ADM} = \int_0 ^\infty \! dR \, 4\pi R^2 \, \rho(R)  \; , 
\ee
where $\rho(R)$ is the local, `rest' energy density, as given by
\be
T_{00} =  g_{00} \, \rho \; .
\ee
At first this is surprising, since eq.~(\ref{integralrho}) only involves the matter energy density, whereas the ADM mass should receive contributions from the gravitational field's stress-energy as well. But those contributions are secretly taken into account by eq.~(\ref{integralrho}), which lacks the $\sqrt{g}$ factor one would need to define the total mass of matter alone.
Now, in our case eq.~(\ref{integralrho}) can be split into an integral inside the source, where both $\varphi$ and the source itself contribute to $\rho$, and an integral outside, where only $\varphi$ contributes.
Even without knowing the solution inside the source, we can use eq.~(\ref{integralrho}) to put a lower bound on the total mass,
\be
M_{\rm ADM} > \int_{R_*} ^\infty \! dR \, 4\pi R^2 \, \rho_\varphi(R)  \; , 
\ee
where $R_*$ is the source's radius in this new radial variable. We are of course assuming that the total $\rho$ inside the source is positive. Since the scalar's energy density would blow up at the singularity at $r=2Gm$, we see that we must keep $r_*$ away from the singularity for the total mass to be finite, as given by eq.~(\ref{mass}).
More precisely, going back to our original radial variable $r$,
\begin{eqnarray}
\rho_{\varphi} & = & \sfrac12 g^{RR} \left(\sfrac{d}{dR} \varphi\right)^2 = \sfrac12 g^{rr} \left(\sfrac{d }{dr} 
\varphi\right)^2 = \sfrac12 \, \varphi_0^2 \, f'^2/f^{2-\beta}\\
R & = & r f(r)^{\frac{1-\beta}{2}} \; ,
\end{eqnarray}
we have
\begin{eqnarray} 
M_{\rm ADM} & > &  \int_{r_*} ^\infty \! dr \, \frac{dR}{dr} \, 4\pi r^2 f^{1-\beta} \, \rho_\varphi(R) \\
			& = &  2 \pi \varphi_0^2  \int_{r_*} ^\infty \! dr \, \frac{r^2 f'^2}{f^{\frac{1+\beta}{2}}} 
			\left[ 1 + \frac{(1-\beta)}{2} \, \frac{rf'}{f}\right] \\
			& = & \beta m + \frac{m}{4} \bigg[ \frac{(1-\beta)^2}{ f_*^\frac{1+\beta}{2}  } -  (1+\beta)^2 \, 
				 f_*^\frac{1-\beta}{2} \bigg]\; ,
\end{eqnarray}
where we used the relation between $\beta$ and $\varphi_0$, eq.~(\ref{beta}), and we defined $f_*$ as the value of $f(r)$ at $r_*$. The ADM mass is $\beta m$ (eq.~(\ref{mass})), so for the inequality to hold the term in brackets must be negative, that is
\be
f_* > \frac{(1-\beta)^2}{(1+\beta)^2 } \; .
\ee
Then the overall field excursion, eq.~(\ref{phistar}), is bounded by
\be
(\Delta \varphi)_{\rm out}^2 < \frac{1}{4 \pi G} \cdot (1 - \beta^2) \log ^2\frac{1+\beta}{1 - \beta} \; ,
\ee
where once again we used that $\varphi_0^2 = (1-\beta^2)/(16 \pi G)$.
Maximizing over $\beta$ we finally get as absolute bound
\be \label{GRbound}
G \cdot (\Delta \varphi)_{\rm out}^2 < \frac{\xi}{4 \pi} = 0.1398... \; ,
\ee
where $\xi = 1.7569...$ is defined as
\be
\xi \equiv \max_{0< \beta <1} \bigg[ (1 - \beta^2) \log ^2\frac{1+\beta}{1 - \beta} \bigg]  \; .
\ee


\section{Discussion}\label{discussion}

We have exhibited a quite generic tension between trying to source and measure a super-Planckian field excursion, and controlling the gravitational backreaction not to make the experiment collapse into a black hole. In the Newtonian, static approximation one can show---at least in the single field case---that the optimal field configurations are spherically symmetric. Then the overall variation a scalar can undergo {\em outside} its source is always sub-Planckian, and given by eq.~(\ref{outside}). {\em Inside} the source there is apparently no absolute bound, and the overall $\Delta \varphi$ can be made large at will, eq.~(\ref{bound}), but at the expense of building exponentially large experiments. Exactly the same conclusion holds in any number of dimensions, and in a multi-field, moduli space-like case, eq.~(\ref{multi}), where the bound applies to the invariant length in field space as measured along radial lines in real space. It is curious that a log-divergence for the maximum field variation also appears in different contexts \cite{swampland1, swampland2}, although we see no direct connection with our logarithmic enhancement of
eq.~(\ref{bound}).

The analysis can be extended to General Relativity. Assuming staticity, spherical symmetry, and positivity of the total energy density, we showed that there is an absolute bound  for $\Delta \varphi$ outside the source of order of the Planck scale, 
\be
G \cdot (\Delta \varphi)_{\rm out}^2 < \frac{\xi}{4 \pi} = 0.1398...
\ee
The question is obviously what happens {\em inside} the source. In particular, in the Newtonian analysis we never worried about the source's internal dynamics, nor about its gravitational backreaction. We just assumed that we can arrange sources for $\varphi$ any way we want, and that they do not lead to any sizable  gravitational field. But in GR the situation is more constrained. For one thing, if the $\varphi$ profile does not solve a free equation, its stress-energy tensor is not conserved by itself. Then the source's stress-energy tensor has to fix the total stress-energy conservation, hence its gravitational backreaction cannot be negligible. Also, one would like to impose that standard energy conditions are obeyed, most notably the dominant/null energy condition \cite{nec}. 
One possibility is that the logarithmic enhancement we found at the Newtonian level survives a more thorough analysis. One should find a well-behaved physical system that can source the needed scalar profile  while keeping the gravitational backreaction under control.
On the other hand, it may be that once the above constraints are taken into account, the optimal configuration we found in the Newtonian case is not viable anymore, and we get an absolute, Planckian bound on the overall field variation we can have inside the source. Or perhaps other effects, like quantum corrections to the $\varphi$ stress-energy tensor coming from graviton loops, are enough to spoil the logarithmic unboundedness we found at the classical level, thus making the maximum allowed $\Delta \varphi$ finite. Also one would like to extend the GR analysis to non-spherically symmetric, non-static configurations.
Robustly concluding in General Relativity that super-Planckian field values are not observable in asymptotically flat space, would certainly raise doubts about their physicality, perhaps suggesting that a theory of quantum gravity should dispose of them.

\section*{Acknowledgments}
I am grateful to Nima Arkani-Hamed, Puneet Batra, Raphael Bousso,  Ben Freivogel, Liam McAllister, and Massimo Porrati for useful discussions and comments.


\begin{thebibliography}{99}

\bibitem{swampland1}
  C.~Vafa,
  ``The string landscape and the swampland,''
  arXiv:hep-th/0509212.
  
\bibitem{swampland2}
H.~Ooguri and C.~Vafa,
  ``On the geometry of the string landscape and the swampland,''
  Nucl.\ Phys.\  B {\bf 766}, 21 (2007)
  [arXiv:hep-th/0605264].

\bibitem{banks}
  T.~Banks, M.~Dine, P.~J.~Fox and E.~Gorbatov,
  ``On the possibility of large axion decay constants,''
  JCAP {\bf 0306}, 001 (2003)
  [arXiv:hep-th/0303252].

\bibitem{liam}
  L.~McAllister and E.~Silverstein,
  ``String Cosmology: A Review,''
  Gen.\ Rel.\ Grav.\  {\bf 40}, 565 (2008)
  [arXiv:0710.2951 [hep-th]].

\bibitem{lyth}
  D.~H.~Lyth,
  ``What would we learn by detecting a gravitational wave signal in the  cosmic
  microwave background anisotropy?,''
  Phys.\ Rev.\ Lett.\  {\bf 78}, 1861 (1997)
  [arXiv:hep-ph/9606387].

\bibitem{paolo}
  N.~Arkani-Hamed, H.~C.~Cheng, P.~Creminelli and L.~Randall,
  ``Extranatural inflation,''
  Phys.\ Rev.\ Lett.\  {\bf 90}, 221302 (2003)
  [arXiv:hep-th/0301218].

\bibitem{weakgravity}
  N.~Arkani-Hamed, L.~Motl, A.~Nicolis and C.~Vafa,
  ``The string landscape, black holes and gravity as the weakest force,''
  JHEP {\bf 0706}, 060 (2007)
  [arXiv:hep-th/0601001].

\bibitem{buchdahl}
  H.~A.~Buchdahl,
  ``Reciprocal Static Metrics and Scalar Fields in the General Theory of
  Relativity,''
  Phys.\ Rev.\  {\bf 115}, 1325 (1959).

\bibitem{wyman}
  M.~Wyman,
  ``Static Spherically Symmetric Scalar Fields In General Relativity,''
  Phys.\ Rev.\  D {\bf 24}, 839 (1981).

\bibitem{nohair}
  J.~D.~Bekenstein,
  ``Nonexistence of baryon number for static black holes,''
  Phys.\ Rev.\  D {\bf 5}, 1239 (1972).


\bibitem{positiveenergy1}
  R.~Schon and S.~T.~Yau,
  ``Proof Of The Positive Mass Theorem. 2,''
  Commun.\ Math.\ Phys.\  {\bf 79}, 231 (1981).
 
 \bibitem{positiveenergy2}
 E.~Witten,
  ``A Simple Proof Of The Positive Energy Theorem,''
  Commun.\ Math.\ Phys.\  {\bf 80}, 381 (1981).

\bibitem{weinberg}
  S.~Weinberg,
  ``Gravitation and Cosmology,''
{\it  John Wiley \& Sons, 1972.}

\bibitem{nec}
  S.~Dubovsky, T.~Gregoire, A.~Nicolis and R.~Rattazzi,
  ``Null energy condition and superluminal propagation,''
  JHEP {\bf 0603}, 025 (2006)
  [arXiv:hep-th/0512260].


\end{thebibliography}
\end{document}